\documentclass[aps,prd,showpacs,preprint,amsmath,amssymb,11pt]{revtex4}
\usepackage{tabularx}
\usepackage{here}
\usepackage{graphicx}
\usepackage{subfigure}
\usepackage{epsfig}
\usepackage{color}
\usepackage{slashed}
\usepackage{hyperref}
\usepackage{float}
\begin{document}

\title{$QQ^{\prime} \bar u \bar d$ bound state in the Bethe-Salpeter equation approach}

\author{G.-Q. Feng$^{1}$\footnote{fenggq@ihep.ac.cn}, X.-H. Guo$^2$\footnote{Corresponding author. xhguo@bnu.edu.cn},
and B.-S. Zou$^{3,1}$\footnote{Corresponding author. zoubs@ihep.ac.cn}}

\affiliation{$^1$Institute of High Energy Physics and Theoretical Physics Center for Science Facilities, Chinese Academy of Sciences, Beijing 100049, China\\
$^2$College of Nuclear Science and Technology, Beijing Normal
University, Beijing 100875, China\\
$^3$State Key Laboratory of Theoretical Physics, Institute of Theoretical Physics,
Chinese Academy of Sciences, Beijing 100190,China}

\begin{abstract}
In the heavy quark limit, we establish the Bethe-Salpeter equations for the ground state $QQ^{\prime} \bar u \bar d$ containing one
heavy diquark $QQ^{\prime}$ ($Q, Q^{\prime}=b$ or $c$) and one light antidiquark $\bar u \bar d$. We solve the Bethe-Salpeter
equations numerically in the covariant instantaneous approximations with the kernels containing a scalar
confinement term and a one-gluon-exchange term. Numerical solutions for the Bethe-Salpeter wave functions are
presented. The results show that the masses of $bb \bar u \bar d$, $bc \bar u \bar d$, and $cc \bar u \bar d$ bound states
lie below the threshold of $\bar{B}^{*0}\,B^-$ or $B^0\,{B^*}^-$, $B^-D^+$ or $\bar{B}^{0} D^0$,
and $D^+\,{D^*}^0$ or ${D^*}^+\,D^0$ mesons, respectively. The ground states $QQ^{\prime} \bar u \bar d$ may exist and we expect the
forthcoming experimental data to confirm them.
\end{abstract}
\maketitle

\section{Introduction}

More and more new heavy hadron states have been measured in recent years.
However, many of these states have special properties, and it is difficult to explain them as ordinary mesons with quark-antiquark ($q \bar q$) structure or baryons with three-quark ($qqq$) structure. This stimulates much enthusiasm from physicists to understand the structure of these states. Various models have been used, i.e. hadronic molecular states \cite{Voloshin;1976}, multi-quark clusters \cite{Wilczek;2003,Zou;2005,An;2006,Liu;2006}, hybrid states \cite{Jaffe;1976}, etc.. The existence of tetraquark hadrons composed of two quarks and two anti-quarks was originally suggested in Refs. \cite{Jaffe;1977,Jaffe;1978} and the states of $QQ^{\prime} \bar u \bar d$ ($Q, Q^{\prime}=b$ or $c$) were studied theoretically in different approaches \cite{Zouzou;1986,Lipkin;1986,Heller;1987,Carlson;1988,Pepin;1997,Brink;1998,Silvestre;1993,Navarra;2007,Zhang;2008,Vijande;2006,
Vijande;2009,Yang;2009,Du;2013}. In Refs. \cite{Lipkin;1986,Pepin;1997,Vijande;2009}, it was pointed out that both $bb\bar u \bar d$ and $cc\bar u \bar d$ can form bound states. In Refs. \cite{Zouzou;1986,Heller;1987,Carlson;1988,Brink;1998,Silvestre;1993,Navarra;2007,Zhang;2008,Vijande;2006,
Yang;2009,Du;2013}, it was found that only $bb\bar u \bar d$ lies below the threshold of two-bottom mesons. It is obvious that the existence and stability of $QQ^{\prime} \bar u \bar d$ are model dependent.

When considering the ground states, two identical quarks can only constitute an axial-vector diquark with spin $1$, while two quarks with
different flavors can constitute either a scalar diquark with spin $0$ or an axial-vector diquark with spin $1$. Thus for the scalar $\bar u \bar d$ system (we do not consider the axial-vector $\bar u \bar d$ system in this paper), the quantum numbers of ground states $bb\bar u \bar d$ and $cc\bar u \bar d$ are $J^P=1^+$, while the quantum numbers of the ground states $bc\bar u \bar d$ can be $J^P=0^+$ and $J^P=1^+$. Theoretically, the physical processes including heavy quarks
can be simplified by using the heavy quark effective theory
(HQET) \cite{hqet1,hqet2,hqet3,hqet4,hqet5,hqet6,hqet7,hqet8,hqet9}. In this paper, in the heavy quark limit, we will try to interpret the ground states $QQ^{\prime} \bar u \bar d$ as bound states of one heavy diquark $QQ^{\prime}$ and one light antidiquark $\bar u \bar d$. We will establish the Bethe-Salpeter (BS) equations \cite{Salpeter;1951,lurie;1968,Zuber;1980} for $bb\bar u \bar d$, $bc\bar u \bar d$, and $bc\bar u \bar d$ bound states, respectively, assuming that the kernels contain a scalar confinement term and a one-gluon-exchange term motivated by the potential model. In order to reflect the fact that the diquarks have finite size, in the kernel we will introduce form factors in the vertex of diquarks coupling to gluon to describe the structure of diquarks \cite{Guo;1996,Guo;1999,Guo;2000,Guo;2001,Guo;2007,Guo;2011}. We will solve BS wave
functions numerically in the covariant instantaneous approximation. The results will show that $bb\bar u \bar d$, $bc\bar u \bar d$, and $cc\bar u \bar d$ bound states may exist with reasonable parameters in our model.

The remaining of this paper is organized as follows. In
Sec. \ref{formalism}, we will establish the BS equations for $bb\bar u \bar d$, $cc\bar u \bar d$, and $bc\bar u \bar d$ bound
states assuming that the kernels contain scalar confinement and one-gluon-exchange terms. In Sec. \ref{results}, we will solve BS wave
functions numerically in the covariant instantaneous approximation and present the numerical solutions of the BS
equations and their dependence on the parameters in our
model. Finally, Sec. \ref{summary} is reserved the summary and discussion.

\section{\label{formalism} BS equations for $QQ^{\prime} \bar u \bar d$}

\subsection{BS equations for $QQ^{\prime} \bar u \bar d$ bound states with $J^P=1^+$}

In this section we will establish the BS
equations for ground states of $bb\bar u \bar d$, $bc\bar u \bar d$, and $cc\bar u \bar d$ with $J^P=1^+$. We start by defining the BS wave function for the bound
state $|QQ^{\prime} \bar u \bar d(P)\rangle$ as
\begin{eqnarray}\label{bsw}
\chi_P^{\mu}(x_{1},x_{2})=\langle0|T \phi_{\bar u \bar d}(x_1)\varphi_{QQ^{\prime}}^{\mu}(x_2)|QQ^{\prime} \bar u \bar d(P)\rangle,
\end{eqnarray}
where $\phi_{\bar u \bar d}(x_{1})$ is the field operator of
the light antidiquark $\bar u \bar d$, and $\varphi_{QQ^{\prime}}^{\mu}(x_2)$ is the field operator of the heavy diquark $QQ^{\prime}$,
$P=M_{QQ^{\prime} \bar u \bar d}\,\,v$ is the momentum of the bound state with mass $M_{QQ^{\prime} \bar u \bar d}$ and
velocity $v$.

Let us define
\begin{eqnarray}
\lambda_1=\frac{m_1}{m_1+m_2},~~~~~~~~~~~~~~~\lambda_2=\frac{m_2}{m_1+m_2},
\end{eqnarray}
with $m_{1}$ and $m_{2}$ being the masses of the $\bar u \bar d$ and $QQ^{\prime}$ diquarks,
respectively, and let $p$ be the relative
momentum between the two constituents. In momentum
space, the BS wave function can be written as
\begin{eqnarray}
\chi_P^{\mu}(x_1,x_2)=e^{-iP\cdot X}\int \frac{d^4
p}{(2\pi)^4}\,e^{-ip\cdot x}\,\chi_P^{\mu}(p),
\end{eqnarray}
where $X$ and $x$ are the center-of-mass coordinate and the relative
coordinate which are, respectively, defined as
$X\equiv\lambda_{1}x_{1}+\lambda_{2}x_{2}, ~x\equiv x_{1}-x_{2}.$

The BS equation for the $QQ^{\prime} \bar u \bar d$ bound state with $J^P=1^+$ can be
written in the following form:
\begin{eqnarray}\label{bsequation}
\chi_{P}^{\mu}(p)=S(p_2)^{\mu \rho}\int \frac{d^4 p^{\prime}}{(2\pi)^4}\,G_{\rho \nu}(P,p,p^{\prime})\, \chi_{P}^{\nu}(p^{\prime})\,S(p_1),
\end{eqnarray}
where $S(p_1)$ and $S(p_2)^{\mu \rho}$ are the propagators of $\bar u \bar d$ and $QQ^{\prime}$ diquarks, respectively, the momenta
of $\bar u \bar d$ and $QQ^{\prime}$ are related to $P$ and $p$ as $p_1=\lambda_1 P+p$,
$p_2=\lambda_2P-p$, $G_{\rho \nu}(P,p,p^{\prime})$ is the interaction kernel
which contains two particle irreducible diagrams.

For convenience, we use the following forms of longitudinal
and transverse relative momenta:
\begin{eqnarray}
p_l=v\cdot p-\lambda_2 M,~~~~~~~~~~~~~~~~p_t=p-(v\cdot p)v.
\end{eqnarray}

The light antidiquark propagator can be written as the following form:
\begin{eqnarray}\label{propergater1}
S(p_1)=\frac{i}{(\lambda_1 M+p_l)^2-w_1^2+i\epsilon},
\end{eqnarray}
where $w_1\equiv\sqrt{p_t^2+m_1^2}$ with $p_t=|p_t|$.

In the leading order of the $1/m_Q$ expansion, the propagator of the axial-vector heavy diquark has the form
\begin{eqnarray}\label{propergater2}
S(p_2)=\frac{-i(g^{\mu \rho}-v^{\mu}v^{\rho})}{2m_2(\lambda_2M-p_l-m_2+i\epsilon)}.
\end{eqnarray}

The interaction kernel contains two terms in the model, a scalar confinement term $V_1$ and a one-gluon-exchange term $V_2$ \cite{Guo;2011},
\begin{eqnarray}\label{kernelbb0}
iG_{\rho \nu}=2m_2\,g_{\rho \nu}I\otimes I V_1-(p_1+p_1^{\prime})_{\mu} \otimes \Gamma_{\rho \nu} ^{\mu}V_2,
\end{eqnarray}
where
\begin{eqnarray}\label{gamm}
\Gamma_{\rho \nu} ^{\mu}=(p^{\mu}_2+p_2^{\prime \mu})\,g_{\rho \nu}\,F_{V1}(Q^2)-(p_{2\rho}\,g_{\nu}^{\mu}+p_{2\nu}^{\prime}\,g_{\rho}^{\mu})\,F_{V2}(Q^2)+p_{2\rho}\,p_{2\nu}^{\prime}\,(p^{\mu}_2+p_2^{\prime \mu})\,F_{V3}(Q^2),
\end{eqnarray}
with $p_1$ and $p_1^{\prime}$ ($p_2$ and $p_2^{\prime}$) being the momenta of the light (heavy) diquark attached to the gluon, and $F_{Vi}\,(i=1,2,3)$ being form factors due to the finite size of the heavy diquark. Since the contribution from the third term in Eq. (\ref{gamm}) is suppressed at small and intermediate momentum transfer \cite{Kroll;1987}, in our calculation, this term is ignored. According to the result in \cite{Guo;2011}, the form factor $F_{V2}(Q^2)=0$.

Consequently, the interaction kernel can be rewritten as the following form:
\begin{eqnarray}\label{kernelbb}
iG_{\rho \nu}=2m_2\,g_{\rho \nu}I\otimes I V_1-g_{\rho \nu}F(Q^2)(p_1+p_1^{\prime})_{\mu}\otimes(p_2+p_2^{\prime})^{\mu}V_2,
\end{eqnarray}
where the notation $F(Q^2)=F_{V1}(Q^2)$ has been used. The form factor $F(Q^2)$ was calculated in Ref. \cite{Guo;2011}.

At the vertex of the gluon with the light diquarks, the form factor $F_1(Q^2)$ is introduced and is parameterized as $F_1(Q^2)=\frac{\alpha_s^{(\rm eff)}Q_1^2}{Q^2+Q_1^2}$, where $Q_1^2$ is a parameter which freezes $F_1(Q^2)$ when $Q^2$ is very small \cite{Kroll;1987}.

Using the covariant instantaneous approximation, $p_l=p_l^{\prime}$, in the kernel, the terms $V_1$ and $V_2$ in $G_{\rho \nu}(P,p,p^{\prime})$ are replaced by $\tilde{V_i}\equiv V_i \mid_{p_l=p^{\prime}_l}$ ($i=1,2$).

In the heavy quark limit, as a result of the $SU(2)_s\otimes
SU(2)_f$ symmetry, the dynamics inside the bound state is controlled
by the configuration of the light degrees of freedom. Then, the BS wave function $\chi_{P}^{\mu}(p)$ has the following form:
\begin{eqnarray}\label{scalarbsfunction}
\chi_{P}^{\mu}(p)=\phi_P(p)\,\varepsilon^{\mu},
\end{eqnarray}
where $\phi_P(p)$ is a scalar function, $\varepsilon^{\mu}$ is the polarization of the bound state which satisfies the relation $\varepsilon\cdot v= 0$.

From Eqs. (\ref{bsequation}), (\ref{propergater1})$-$(\ref{scalarbsfunction}), we obtain the following form of the BS equation:
\begin{eqnarray}\label{bseform}
\phi_P(p)&=&\frac{-i}{(\lambda_2M-p_l-m_2+i\epsilon)
[(\lambda_1M+p_l)^2-w_1^2+i\epsilon]}\nonumber\\
&&\times\int\frac{d^3p^{\prime}_t}{(2\pi)^3}\,[\tilde{V}_1-2(\lambda_1M+p_l)
\,F(Q^2)\,\tilde{V}_2]\,\tilde{\phi}_P(p_t^{\prime}),
\end{eqnarray}
where we have used the definition $\tilde{\phi}_P(p_t^{\prime})\equiv \int(dp_l^{\prime}/2\pi)\,\phi_P(p^{\prime})$.

After performing the integration $\int(dp_l/2\pi)$ and applying the residue theorem, the BS equation
for $\tilde{\phi}_P(p_t)$ becomes the following form:
\begin{eqnarray}\label{bsfinall}
\tilde{\phi}_P(p_t)&=&\frac{-1}{2w_1(M-w_1-m_2)}\int\frac{d^3p^{\prime}_t}{(2\pi)^3}\,[\tilde{V}_1-2w_1F(Q^2)
\,\tilde{V}_2]\,\tilde{\phi}_P(p_t^{\prime}),
\end{eqnarray}
with $\tilde{V}_1$ and $\tilde{V}_2$ having the forms
\begin{eqnarray}\label{v1}
\tilde{V}_1=\frac{8\pi\kappa}{[(p_t-q_t)^2+\mu^2]^2}-(2\pi)^3
\delta^3  (p_t-q_t)
	\int \frac{d^3 k}{(2\pi)^3}\frac{8\pi\kappa}{(k^2+\mu^2)^2},
\end{eqnarray}
\begin{eqnarray}\label{v2}
\tilde{V}_2=-\frac{16\pi}{3}
	\frac{\alpha_{s}^{({\rm eff})2}
Q_{1}^{2}}{[(p_t-q_t)^2+\mu^2][(p_t-q_t)^2+Q_{1}^{2}]},
\end{eqnarray}
where $\kappa$ and $\alpha_s^{(\rm eff)}$ are coupling parameters related to scalar confinement and the
one-gluon-exchange diagram, respectively. The second term in Eq. (\ref{v1}) is the counter term which removes
the infra-red divergence in the integral equation. The parameter $\mu$ is introduced to avoid the infra-red divergence in numerical
calculations. The limit $\mu\rightarrow 0$ is taken at the end of the calculation. The parameter $Q_1^2$ is taken as $Q_1^2=3.2$ GeV$^2$ \cite{Kroll;1987}, the parameters $\kappa$ and $\alpha_s^{(\rm eff)}$ should be related to each other when we solve the eigenvalue equation with a fixed eigenvalue (with the Gauss quadrature rule the BS integral equation is changed into an eigenvalue equation). The parameter $\kappa$ varies in the region between 0.02 GeV$^3$ and 0.1 GeV$^3$ \cite{Guo;1996,Guo;1999,Guo;2000,Guo;2001,Guo;2007,Guo;2011}.

In general, the normalization condition of the BS wave function for the axial-vector bound states $QQ^{\prime} \bar u \bar d$ can be written as \cite{lurie;1968}
\begin{eqnarray}
i\int \frac{d^4p d^4p^{\prime}}{(2\pi)^2}\bar{\chi}_{\mu,P}(p)\bigg[\frac{\partial}{\partial P^0} I^{\mu\nu}(p,p^{\prime})\bigg]\bar{\chi}_{\nu,P}(p^{\prime})=1,
\end{eqnarray}
where $I^{\mu\nu}(p,p^{\prime})$ is the inverse of the four point propagator which has the following form:
\begin{eqnarray}
I^{\mu\nu}(p,p^{\prime})=\frac{1}{3}(2\pi)^4 \delta^4(p-p^{\prime})S^{-1}(p_1)S^{-1,\,\mu\nu}(p_2).
\end{eqnarray}
After integrating out the longitudinal momentum $p_l$, the normalization
condition takes the following form:
\begin{eqnarray}\label{normbb}
&&\frac{2}{3}\lambda_1\int\frac{d^3p_t}{(2\pi)^3}\frac{m_2(\tilde{\phi}_1-2w_1\tilde{\phi}_2)[\tilde{\phi}_1-2(2M-2m_2-w_1)\tilde{\phi}_2]}{2w_1(M-m_2-w_1)^2}\nonumber\\
&&+\frac{2}{3}\lambda_2\int\frac{d^3p_t}{(2\pi)^3}\frac{(M-w_1)(\tilde{\phi}_1-2w_1\tilde{\phi}_2)^2}{2w_1(M-m_2-w_1)^2}=1,
\end{eqnarray}
where $\tilde{\phi}_1=\int d^3p^{\prime}_t/(2\pi)^3\,\tilde{V}_1\,\tilde{\phi}_P(p_t^{\prime})$ and $\tilde{\phi}_2=\int d^3p^{\prime}_t/(2\pi)^3\,\tilde{V}_2\,\tilde{\phi}_P(p_t^{\prime})$.

\subsection{BS equation for $bc \bar u \bar d$ bound states with $J^P=0^+$}

As pointed out in Introduction, two quarks with different flavors can constitute either a scalar diquark or an axial-vector diquark. In this section we will establish the BS equation for the ground state of $bc\bar u \bar d$ with $J^P=0^+$. We can define the BS wave function of the bound state $|bc \bar u \bar d(P)\rangle$ as
\begin{eqnarray}\label{bswbc}
\chi_P(x_{1},x_{2})=\langle0|T \phi_{\bar u \bar d}(x_1)\phi_{bc}(x_2)|bc \bar u \bar d(P)\rangle,
\end{eqnarray}
where $\phi_{bc}(x_2)$ is the field operator of the heavy scalar diquark $bc$.

In momentum space, the BS equation for the $bc\bar u \bar d$ bound state with $J^P=0^+$ can be
written in the following form:
\begin{eqnarray}\label{bsequationbc}
\chi_{P}(p)=S(p_2)\int \frac{d^4 p^{\prime}}{(2\pi)^4}\,G(P,p,p^{\prime})\, \chi_{P}(p^{\prime})\,S(p_1),
\end{eqnarray}
where $S(p_2)$ and $S(p_1)$ are propagators of the heavy diquark and the light antidiquark, respectively.

In the leading order of the $1/m_Q$ expansion, the propagator of the scalar heavy diquark has the form
\begin{eqnarray}\label{propergaterbc}
S(p_2)=\frac{i}{2m_2(\lambda_2 M-p_l-m_2+i\epsilon)}.
\end{eqnarray}

Similar to the case of the axial-vector bound state $QQ^{\prime} \bar u \bar d$, we assume that the kernel also contains two terms in the model, a scalar confinement term $V_1$ and a one-gluon-exchange term $V_2$,
\begin{eqnarray}\label{kernelbc}
-i\,G=2m_2\,I\otimes I V_1-F_{S}(Q^2)(p_1+p_1^{\prime})_{\mu}\otimes(p_2+p_2^{\prime})^{\mu}V_2,
\end{eqnarray}
where $F_{S}$ is the form factor at the vertex of the gluon with the scalar heavy diquark. From the conclusion of Ref. \cite{Guo;2011}, we can see that the form factors for both the effective
vertex of the scalar heavy diquark coupling to the gluon
and the effective vertex of the axial-vector heavy diquark
coupling to the gluon are equal to each other in the leading
order of the $1/m_{Q}$ expansion, $F_{V1}=F_{S}=F(Q^2)$.

After substituting Eqs. (\ref{propergater1}), (\ref{propergaterbc}), and (\ref{kernelbc}) into Eq. (\ref{bsequationbc}) and integrating out the
longitudinal momenta $p_l$ and $p_l^{\prime}$, the BS equation takes the form
\begin{eqnarray}\label{bsfinallbc}
\tilde{\chi}_P(p_t)&=&\frac{-1}{2w_1(M-w_1-m_2)}\int\frac{d^3p^{\prime}_t}{(2\pi)^3}\,[\tilde{V}_1-2w_1F(Q^2)
\,\tilde{V}_2]\,\tilde{\chi}_P(p_t^{\prime}),
\end{eqnarray}
where $\tilde{V}_1$ and $\tilde{V}_2$ have the forms in Eqs. (\ref{v1}) and (\ref{v2}), respectively.

Analogously, the normalization condition of the BS wave function for the scalar bound state $bc \bar u \bar d$ can be written as
\begin{eqnarray}
i\int \frac{d^4p d^4p^{\prime}}{(2\pi)^2}\bar{\chi}_{P}(p)\bigg[\frac{\partial}{\partial P^0} I(p,p^{\prime})\bigg]\bar{\chi}_{P}(p^{\prime})=1,
\end{eqnarray}
with $I(p,p^{\prime})=\frac{1}{3}(2\pi)^4 \delta^4(p-p^{\prime})S^{-1}(p_1)S^{-1}(p_2)$ being the inverse of the four point propagator.

After integrating out the longitudinal momentum $p_l$, the normalization condition for the scalar $bc \bar u \bar d$ bound state has the form
\begin{eqnarray}\label{normbc}
&&\frac{2}{3}\lambda_1\int\frac{d^3p_t}{(2\pi)^3}\frac{m_2(\tilde{\phi}_1-2w_1\tilde{\phi}_2)[\tilde{\phi}_1-2(2M-2m_2-w_1)\tilde{\phi}_2]}{2w_1(M-m_2-w_1)^2}\nonumber\\
&&+\frac{2}{3}\lambda_2\int\frac{d^3p_t}{(2\pi)^3}\frac{(M-w_1)(\tilde{\phi}_1-2w_1\tilde{\phi}_2)^2}{2w_1(M-m_2-w_1)^2}=1.
\end{eqnarray}
We can see that the final forms of BS equations (Eqs.(\ref{bsfinall}) and (\ref{bsfinallbc})) and the normalization conditions (Eqs.(\ref{normbb}) and (\ref{normbc})) for the axial-vector bound state $QQ^{\prime} \bar u \bar d$ and the scalar bound state $bc\bar u \bar d$ are the same. This is because in the heavy quark limit the internal structure is blind to the flavor and spin direction of the heavy constituent and is controlled by the light degrees of freedom, the $\bar u\bar d$ light diquark.

\section{\label{results}The numerical results}

\begin{table}
\caption{Values  of  $\kappa$  and  $\alpha_{s}^{({\rm eff})}$
for different values of $M_{bb\bar u\bar d}(J^P=1^+)$}
\label{bbt}
\begin{center}
\begin{tabular}{lccccc}
\hline
\hline
M(GeV) & $M_{bb\bar u\bar d}=10.60$ & &$M_{bb}=9.90$  & &$M_{\bar u\bar d}=0.80$\\
\hline
$\kappa$(GeV$^3$)&0.02&0.04&0.06 &0.08 &0.10\\
\hline
$\alpha_{s}^{({\rm eff})}$ &0.677&0.709&0.734&0.754&0.771\\
\hline
\hline
M(GeV) & $M_{bb\bar u\bar d}=10.50$ & &$M_{bb}=9.80$  & &$M_{\bar u\bar d}=0.75$\\
\hline
$\kappa$(GeV$^3$)&0.02&0.04&0.06 &0.08 &0.10\\
\hline
$\alpha_{s}^{({\rm eff})}$ &0.620&0.665&0.696&0.721&0.742\\
\hline
\hline
M(GeV) & $M_{bb\bar u\bar d}=10.39$ & &$M_{bb}=9.75$  & &$M_{\bar u\bar d}=0.65$\\
\hline
$\kappa$(GeV$^3$)&0.02&0.04&0.06 &0.08 &0.10\\
\hline
$\alpha_{s}^{({\rm eff})}$ &0.584&0.647&0.686&0.716&0.738\\
\hline
\hline
$\bar{B}^{*0}\,B^-$ or $B^0\,{B^*}^-$& threshold (GeV) & & & 10.605 & \\
\hline
\hline
\end{tabular}
\end{center}
\end{table}

\begin{figure}
\begin{minipage}{\columnwidth}
\centering
\includegraphics[width=8.9cm]{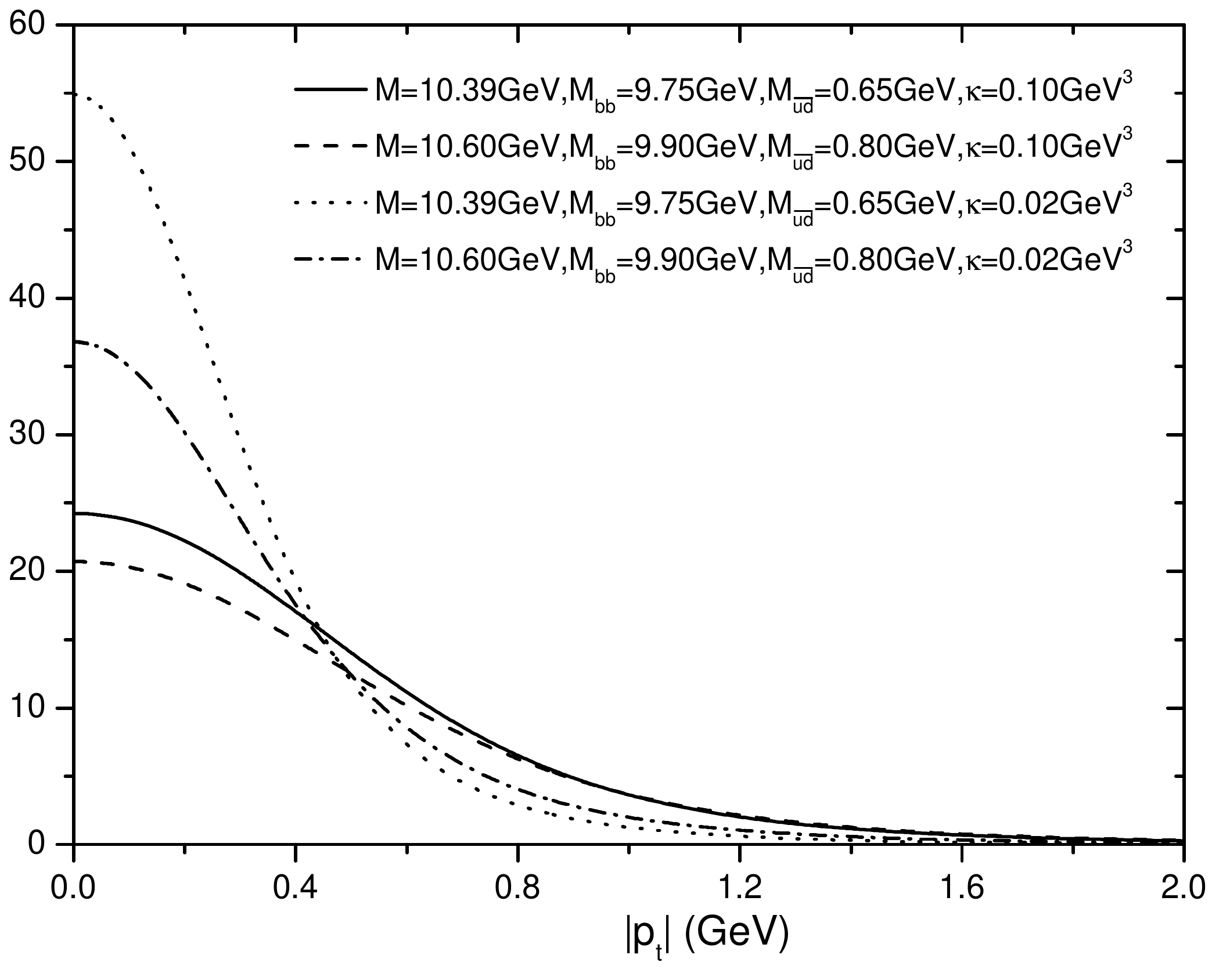}
\end{minipage}
\caption{\small The normalized BS scalar wave functions for $bb\bar u\bar d$. For $\kappa=0.10$ GeV$^3$ and $\kappa=0.02$ GeV$^3$, we show the dependence on $|p_t|$ for two values of $M_{\bar u\bar d}$ and $M_{bb}$. The solid and dotted lines are for $M_{\bar u\bar d}=0.65$ GeV and $M_{bb}=9.75$ GeV, the dashed and dashdotted lines are for $M_{\bar u\bar d}=0.80$ GeV and $M_{bb}=9.90$ GeV, respectively.}
\label{bb1}
\end{figure}

\begin{figure}
\begin{minipage}{\columnwidth}
\centering
\includegraphics[width=8.9cm]{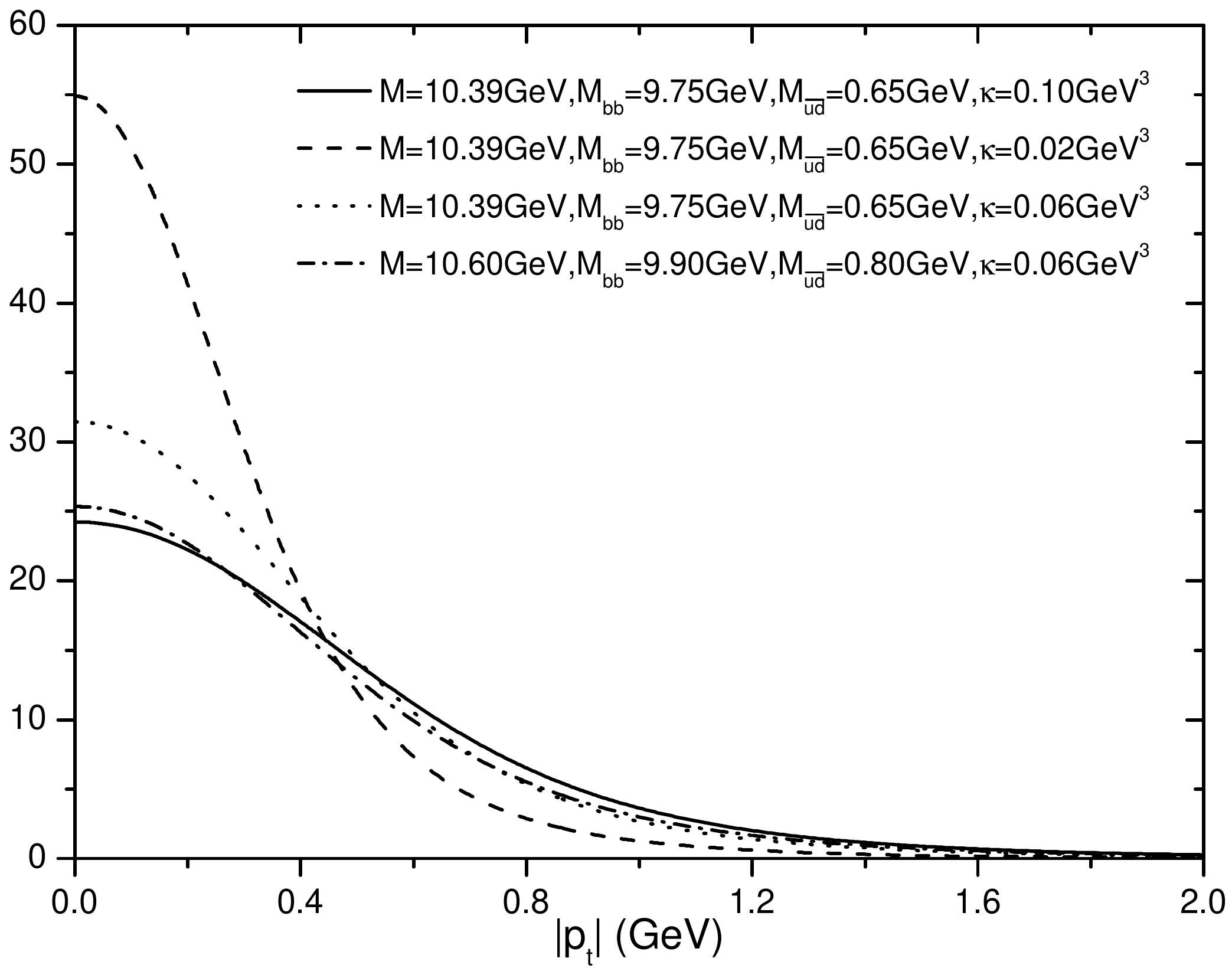}
\end{minipage}
\caption{\small The normalized BS scalar wave functions for $bb\bar u\bar d$. For $M_{\bar u\bar d}=0.65$ GeV and $M_{bb}=9.75$ GeV, we show the dependence on $|p_t|$ for three values of $\kappa$. The solid, dotted, and dashed lines are for $\kappa=0.10$ GeV$^3$, $\kappa=0.06$ GeV$^3$, and $\kappa=0.02$ GeV$^3$, respectively. For comparison, the dashdotted line is for $M_{\bar u\bar d}=0.80$ GeV and $M_{bb}=9.90$ GeV with $\kappa=0.06$ GeV$^3$.}
\label{bb2}
\end{figure}

\begin{table}
\caption{Values  of  $\kappa$  and  $\alpha_{s}^{({\rm eff})}$
for different values of $M_{bc\bar u\bar d}(J^P=0^+(1^+))$}
\label{bct}
\begin{center}
\begin{tabular}{lccccc}
\hline
\hline
M(GeV) & $M_{bc\bar u\bar d}=7.14$ & &$M_{bc}=6.44$  & &$M_{\bar u\bar d}=0.80$\\
\hline
$\kappa$(GeV$^3$)&0.02&0.04&0.06 &0.08 &0.10\\
\hline
$\alpha_{s}^{({\rm eff})}$ &0.688&0.722&0.748&0.770&0.788\\
\hline
\hline
M(GeV) & $M_{bc\bar u\bar d}=7.05$ & &$M_{bc}=6.35$  & &$M_{\bar u\bar d}=0.75$\\
\hline
$\kappa$(GeV$^3$)&0.02&0.04&0.06 &0.08 &0.10\\
\hline
$\alpha_{s}^{({\rm eff})}$ &0.628&0.675&0.709&0.735&0.757\\
\hline
\hline
M(GeV) & $M_{bc\bar u\bar d}=6.93$ & &$M_{bc}=6.29$  & &$M_{\bar u\bar d}=0.65$\\
\hline
$\kappa$(GeV$^3$)&0.02&0.04&0.06 &0.08 &0.10\\
\hline
$\alpha_{s}^{({\rm eff})}$ &0.591&0.655&0.697&0.728&0.753\\
\hline
\hline
$B^-D^+$ or $\bar{B}^{0} D^0$& threshold (GeV) & & & 7.144$-$ 7.149& \\
\hline
\hline
\end{tabular}
\end{center}
\end{table}

\begin{figure}
\begin{minipage}{\columnwidth}
\centering
\includegraphics[width=8.9cm]{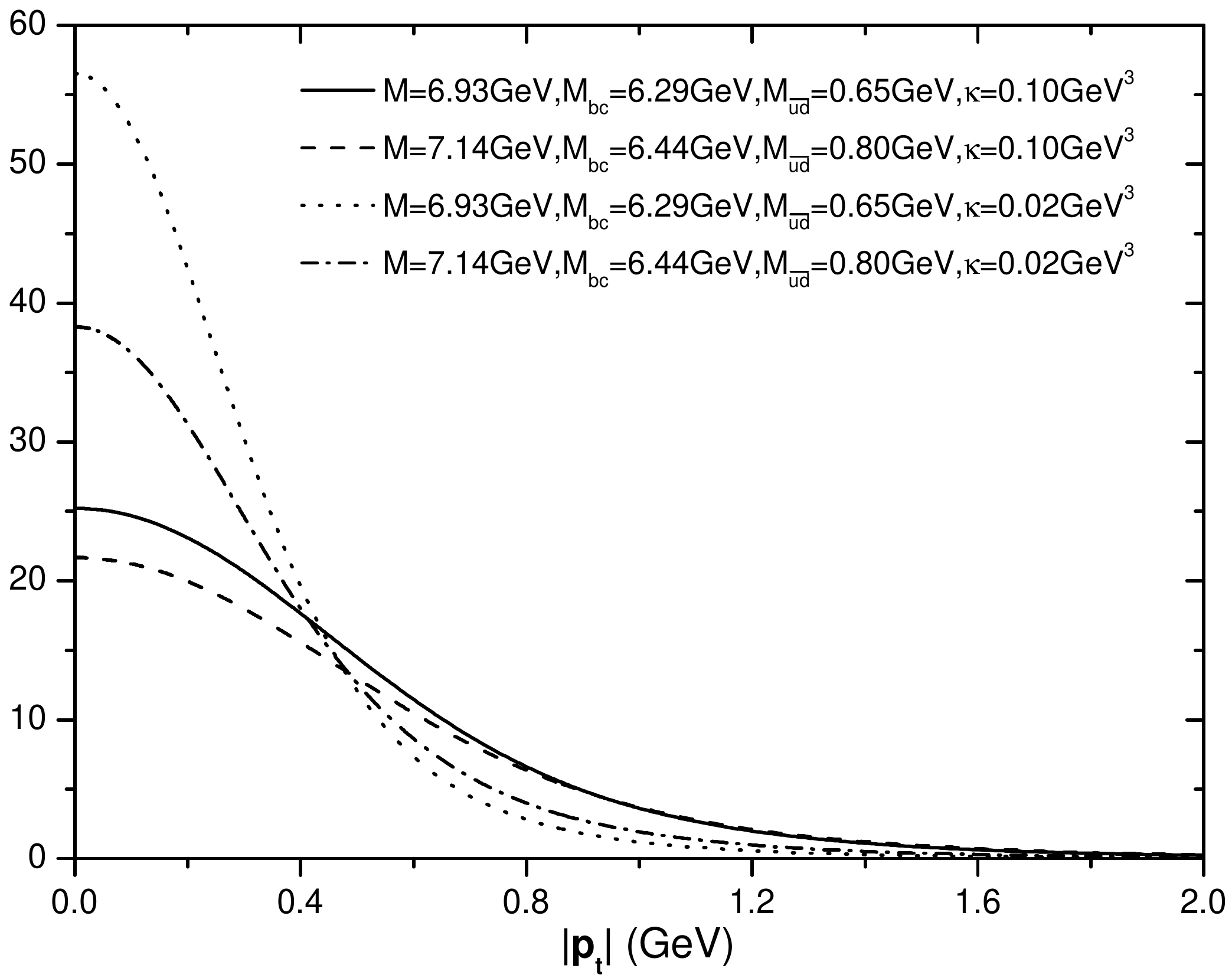}
\end{minipage}
\caption{\small The normalized BS scalar wave functions for $bc\bar u\bar d$. For $\kappa=0.10$ GeV$^3$ and $\kappa=0.02$ GeV$^3$, we show the dependence on $|p_t|$ for two values of $M_{\bar u\bar d}$ and $M_{bc}$. The solid and dotted lines are for $M_{\bar u\bar d}=0.65$ GeV and $M_{bc}=6.29$ GeV, the dashed and dashdotted lines are for $M_{\bar u\bar d}=0.80$ GeV and $M_{bc}=6.44$ GeV, respectively.}
\label{bc1}
\end{figure}

\begin{figure}
\begin{minipage}{\columnwidth}
\centering
\includegraphics[width=8.9cm]{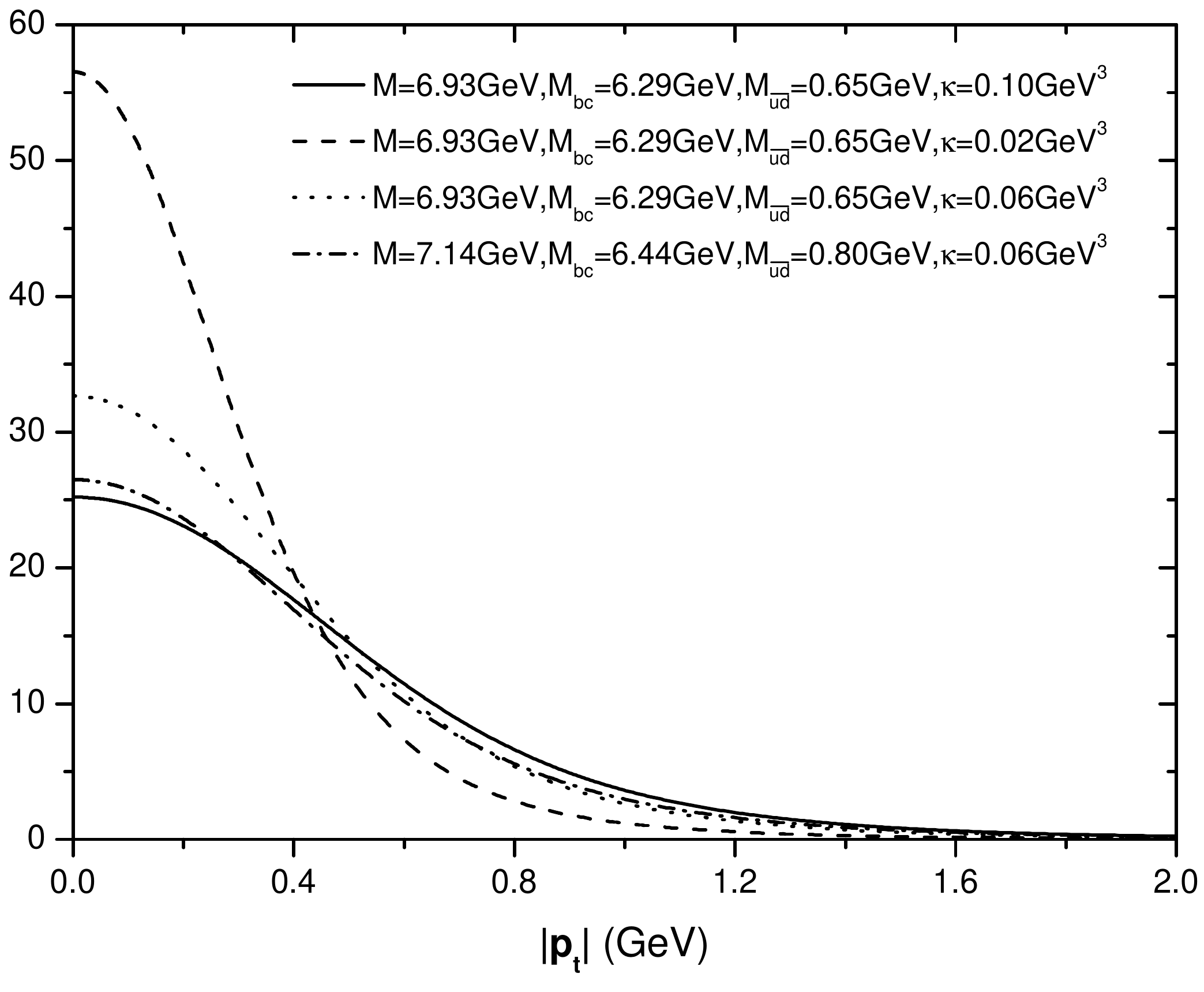}
\end{minipage}
\caption{\small The normalized BS scalar wave functions for $bc\bar u\bar d$. For $M_{\bar u\bar d}=0.65$ GeV and $M_{bc}=6.29$ GeV, we show the dependence on $|p_t|$ for three values of $\kappa$. The solid, dotted, and dashed lines are for $\kappa=0.10$ GeV$^3$, $\kappa=0.06$ GeV$^3$, and $\kappa=0.02$ GeV$^3$, respectively. For comparison, the dashdotted line is for $M_{\bar u\bar d}=0.80$ GeV and $M_{bc}=6.44$ GeV with $\kappa=0.06$ GeV$^3$.}
\label{bc2}
\end{figure}

\begin{table}
\caption{Values  of  $\kappa$  and  $\alpha_{s}^{({\rm eff})}$
for different values of $M_{cc\bar u\bar d}(J^P=1^+)$}
\label{cct}
\begin{center}
\begin{tabular}{lccccc}
\hline
\hline
M(GeV) & $M_{cc\bar u\bar d}=3.87$ & &$M_{cc}=3.17$  & &$M_{\bar u\bar d}=0.80$\\
\hline
$\kappa$(GeV$^3$)&0.02&0.04&0.06 &0.08 &0.10\\
\hline
$\alpha_{s}^{({\rm eff})}$ &0.708&0.746&0.776&0.799&0.820\\
\hline
\hline
M(GeV) & $M_{cc\bar u\bar d}=3.80$ & &$M_{cc}=3.10$  & &$M_{\bar u\bar d}=0.75$\\
\hline
$\kappa$(GeV$^3$)&0.02&0.04&0.06 &0.08 &0.10\\
\hline
$\alpha_{s}^{({\rm eff})}$ &0.645&0.697&0.734&0.763&0.788\\
\hline
\hline
M(GeV) & $M_{cc\bar u\bar d}=3.66$ & &$M_{cc}=3.02$  & &$M_{\bar u\bar d}=0.65$\\
\hline
$\kappa$(GeV$^3$)&0.02&0.04&0.06 &0.08 &0.10\\
\hline
$\alpha_{s}^{({\rm eff})}$ &0.605&0.674&0.720&0.754&0.783\\
\hline
\hline
$D^+\,{D^*}^0$ or ${D^*}^+\,D^0$& threshold (GeV) & & & 3.875$-$ 3.877& \\
\hline
\hline
\end{tabular}
\end{center}
\end{table}

\begin{figure}
\begin{minipage}{\columnwidth}
\centering
\includegraphics[width=8.9cm]{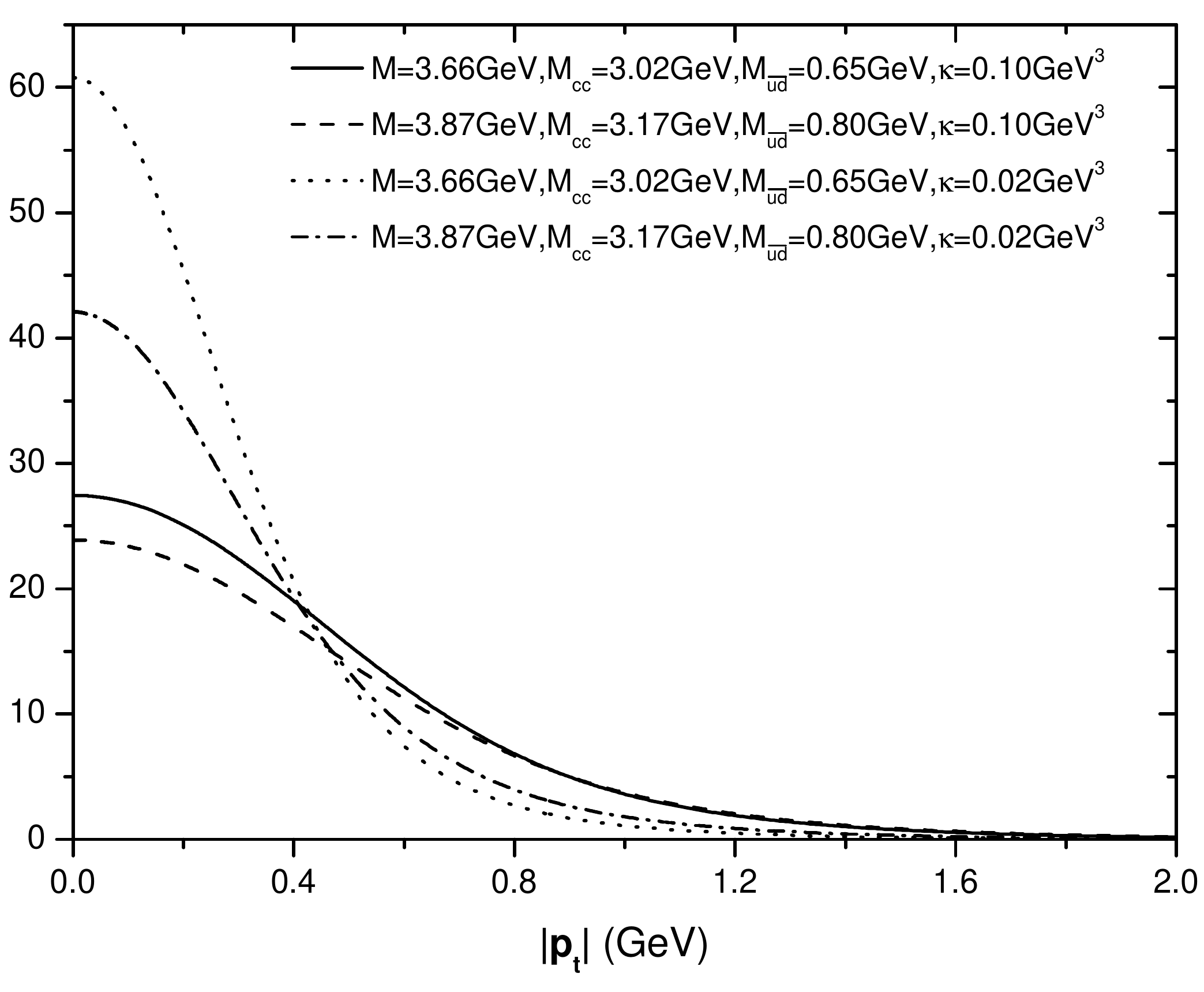}
\end{minipage}
\caption{\small The normalized BS scalar wave functions for $cc\bar u\bar d$. For $\kappa=0.10$ GeV$^3$ and $\kappa=0.02$ GeV$^3$, we show the dependence on $|p_t|$ for two values of $M_{\bar u\bar d}$ and $M_{cc}$. The solid and dotted lines are for $M_{\bar u\bar d}=0.65$ GeV and $M_{cc}=3.02$ GeV, the dashed and dashdotted lines are for $M_{\bar u\bar d}=0.80$ GeV and $M_{cc}=3.17$ GeV, respectively.}
\label{cc1}
\end{figure}

\begin{figure}
\begin{minipage}{\columnwidth}
\centering
\includegraphics[width=8.9cm]{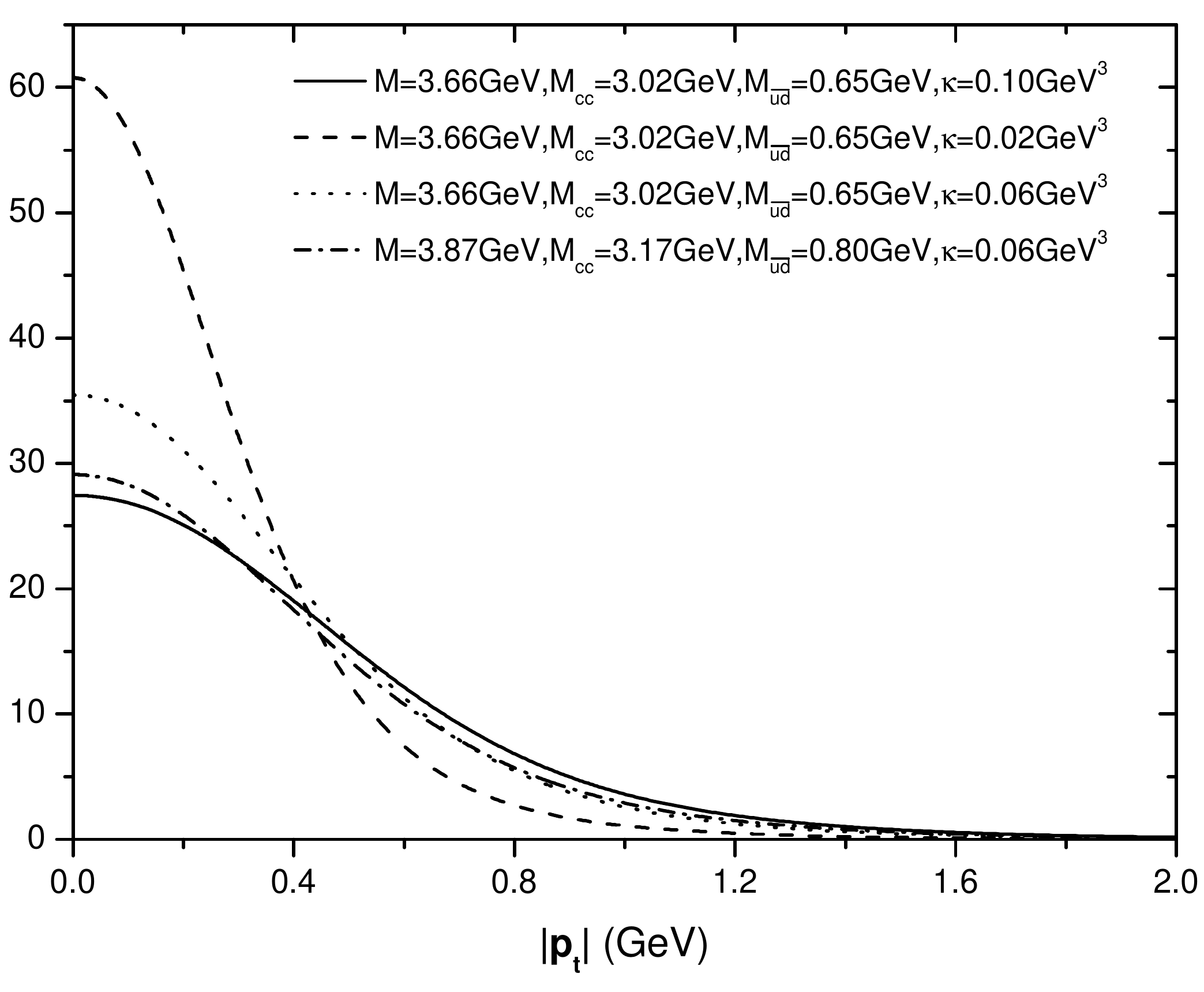}
\end{minipage}
\caption{\small The normalized BS scalar wave functions for $cc\bar u\bar d$. For $M_{\bar u\bar d}=0.65$ GeV and $M_{cc}=3.02$ GeV, we show the dependence on $|p_t|$ for three values of $\kappa$. The solid, dotted, and dashed lines are for $\kappa=0.10$ GeV$^3$, $\kappa=0.06$ GeV$^3$, and $\kappa=0.02$ GeV$^3$, respectively. For comparison, the dashdotted line is for $M_{\bar u\bar d}=0.80$ GeV and $M_{cc}=3.17$ GeV with $\kappa=0.06$ GeV$^3$.}
\label{cc2}
\end{figure}

In our calculation we have several parameters, $\kappa$, $\alpha_{s}^{({\rm eff})}$, $Q_1^2$, $M_{\bar u \bar d}$, $M_{QQ^{\prime}}$, and $M_{QQ^{\prime}}\bar u \bar d$. The binding energy satisfies the relation: $M_{QQ^{\prime}}=M_{QQ^{\prime}}+M_{\bar u \bar d}+E_0$.
The parameter $Q_1^2=3.2$ GeV$^2$ \cite{Kroll;1987}. When we solve the eigenvalue equation with eigenvalue 1, the parameters $\kappa$ and $\alpha_s^{(\rm eff)}$ are related to each other. As discussed in Refs. \cite{Guo;1996,GuoMuta;1996} $\kappa$ is related to $\kappa^{\prime}$ ($\kappa^{\prime}$ is
the confinement parameter in the heavy meson case and is
about $0.2$ GeV$^2$ \cite{Eichten;1978,jinhy;1992,Dai;1993}). Since $\Lambda_{QCD}$ is the only
parameter which is related to confinement, we expect that $\kappa \sim \Lambda_{QCD} \kappa^{\prime}$. In numerical calculation, we let $\kappa$ vary in the region between $0.02$ GeV$^3$ and $0.1$ GeV$^3$. As in Ref. \cite{Guo;1996} the mass of $\bar u \bar d$ is chosen to vary from $0.65$ GeV to $0.80$ GeV. In Ref. \cite{Guo;2011}, the authors solved the BS equations of heavy diquarks and obtained the masses of heavy diquarks $bb$, $bc$, and $cc$. In this paper, we let the mass of the diquark $bb$ vary from $9.65$ GeV to $9.95$ GeV, $bc$ vary from $6.29$ GeV to $6.59$ GeV (the scalar $bc$ diquark and the axial-vector $bc$ diquark have the same masses \cite{Guo;2011,Lozano;1995,White;1991}), and $cc$ vary from $3.02$ GeV to $3.32$ GeV \cite{Guo;2011}. The binding energy $E_0$ varies from around $-10$ MeV to $-100$ MeV. If bound states $M_{QQ^{\prime}\bar u \bar d}$ are stable, the masses of $M_{QQ^{\prime}\bar u \bar d}$ should lie below the threshold of $Q\bar u+Q^{\prime}\bar d$ and $Q^{\prime}\bar u+Q\bar d$ mesons ($M_{QQ^{\prime}\bar u\bar d}< M_{Q \bar u}+M_{Q^{\prime} \bar d}$ and $M_{QQ^{\prime}\bar u\bar d}< M_{Q \bar d}+M_{Q^{\prime} \bar u}$ ($M_{Q \bar u}$ is the mass of the $Q \bar u$ meson, for example)). We let the mass of the bound state $bb\bar u \bar d$ vary from $10.39$ GeV to $10.60$ GeV , $bc\bar u \bar d$ vary from $6.93$ GeV to $7.14$ GeV , and $cc\bar u \bar d$ vary from $3.66$ GeV to $3.87$ GeV, respectively. We vary $\alpha_{s}^{({\rm eff})}$ to find the solutions of BS equations. For each combination of values of $\kappa$, $M_{\bar u \bar d}$, $M_{QQ^{\prime}}$, and $M_{QQ^{\prime}\bar u \bar d}$, $\alpha_{s}^{({\rm eff})}$ takes a certain value.

We present some numerical results in Table \ref{bbt}, Table \ref{bct}, and Table \ref{cct} for $bb\bar u \bar d$, $bc\bar u \bar d$, and $cc\bar u \bar d$ bound states, respectively. According to Refs. \cite{Guo;2011,Lozano;1995,White;1991}, in the leading order of the $1/m_Q$ expansion, the heavy diquark masses are independent of the heavy diquark spin, so we take the same masses for spin 0 and spin 1 $bc\bar u\bar d$ bound states. In Table \ref{bct}, we only list the lower threshold of $B^-D^+$ or $\bar{B}^{0} D^0$ mesons. In Figs. (\ref{bb1}) $-$ (\ref{cc2}), the normalized BS wave functions of $QQ^{\prime}\bar u \bar d$ bound state are shown. It can be seen that the ground states $bb\bar u \bar d$, $bc\bar u \bar d$, and $cc\bar u \bar d$ exist if the parameters take the values in our paper. We have checked that when the binding energy varies between $-10$ MeV and $-100$ MeV, for other values of $M_{QQ^{\prime}\bar u \bar d}$, $M_{QQ^{\prime}}$, and $M_{\bar u \bar d}$, the ground states $QQ^{\prime}\bar u \bar d$ also exist.

\section{\label{summary}Summary and conclusion}

In this work we study ground states $bb \bar u \bar d$, $bc \bar u \bar d$, and $cc \bar u \bar d$ in the BS equation formalism in the heavy quark limit. When $M_Q\rightarrow \infty$, the light degrees of freedom
in a heavy state have good spin and isospin quantum
numbers and the internal structure is blind to the flavor
and spin direction of the heavy system. Based on the picture that these states contain one heavy diquark $QQ^{\prime}$ and one light antiquark $\bar u \bar d$, we establish BS equations for ground states of $bb \bar u \bar d$, $bc \bar u \bar d$, and $cc \bar u \bar d$, respectively. We assume that the kernel contains a scalar confinement term and a one-gluon-exchange term motivated by the potential model. The vertex of the gluon with the diquarks depend on the structure of the diquarks. The form factors are introduced to
describe the internal structure of the diquarks. We solve these BS wave functions numerically in the covariant instantaneous approximation.

In our model, there are several parameters, i.e. $\kappa$, $\alpha_{s}^{({\rm eff})}$, $Q_1^2$, $M_{\bar u \bar d}$, $M_{QQ^{\prime}}$, and $M_{QQ^{\prime}}\bar u \bar d$. In the numerical calculation, we let these parameters vary in some reasonable ranges. As discussed in Sec. \ref{results}, we let $\kappa$ vary in the region between $0.02$ GeV$^3$ and $0.1$ GeV$^3$, and choose $M_{\bar u \bar d}$ to be in the range $0.65$ GeV $-$ $0.80$ GeV. The mass of heavy diquark $bb$ varies from $9.65$ GeV to $9.95$ GeV, $bc$ varies from $6.29$ GeV to $6.59$ GeV, and $cc$ varies from $3.02$ GeV to $3.32$ GeV. The results show that when the binding energy varies between $-10$ MeV and $-100$ MeV, the ground states of $bb \bar u \bar d$, $bc \bar u \bar d$, and $cc \bar u \bar d$ exist when they lie below the threshold of $\bar{B}^{*0}\,B^-$ or $B^0\,{B^*}^-$, $B^-D^+$ or $\bar{B}^{0} D^0$, and $D^+\,{D^*}^0$ or ${D^*}^+\,D^0$ mesons, respectively.
We expect that more experimental data will be collected in future experiments to confirm the existence of these bound states.

\begin{acknowledgements}
This work is supported by the National Natural Science Foundation of China under Grant 10975018, 11175020, 11275025, 11035006, 11121092, 11261130311 (CRC110 by DFG and NSFC), the Chinese Academy of Sciences under Project No. KJCX2-EW-N01, and the Fundamental Research Funds for the Central Universities in China.
\end{acknowledgements}

\end{document}